\documentclass[aps,twocolumn,showpacs,preprintnumbers,superscriptaddress,floatfix,nofootinbib]{revtex4-1}

\usepackage{multirow}
\usepackage{amsfonts}
\usepackage{graphicx,,booktabs}
\usepackage{amssymb}
\usepackage{amsmath,txfonts,ulem}
\usepackage{graphicx}
\usepackage{dcolumn}
\usepackage{color}
\usepackage{bm}
\usepackage{ulem}
\usepackage[colorlinks,
citecolor=blue,
anchorcolor=green,
menucolor=orange,
linkcolor=red,
filecolor=red,
runcolor=pink,
urlcolor=blue,
frenchlinks=red]{hyperref}

\allowdisplaybreaks[4]

\begin{document}

	\title{\boldmath Study on dynamic model of multi-particle spring system}
	
	\author{Xiao-Yun Wang}
	\email{xywang@lut.edu.cn}
	\affiliation{Department of physics, Lanzhou University of Technology,
		Lanzhou 730050, China}
	\affiliation{Lanzhou Center for Theoretical Physics, Key Laboratory of Theoretical Physics of Gansu Province, Lanzhou University, Lanzhou, Gansu 730000, China}
	\author{Jiyuan Zhang}
	\email{jyuanzhang@yeah.net}
	\affiliation{Department of physics, Lanzhou University of Technology,
		Lanzhou 730050, China}
	
	\begin{abstract}
		Generally, multi-particle spring system is an important and widely used physical model. However, with the increase of the number of particles, the difficulty of solving the kinematic trajectory of the particles becomes more and more difficult. The key to solving this problem lies in whether it is possible to construct a dynamic model of the multi-particle spring system which is convenient for numerical solution. In this work, by defining a new physical quantity, the spring action $\mathcal{K}$, we study and reconstruct the dynamic models of two common types of multi-particle spring systems. The calculation results show that the multi-particle spring dynamic model constructed by us has obvious advantages in calculating the spring system with a large number of particles. This will provide important theoretical solutions for the application of multi-particle spring systems to engineering or scientific problems.
	\end{abstract}
	
	
\maketitle
	
	\section{Introduction}
	As one of the basic physical models, the spring multi-particle model is widely used in various life scenarios and engineering fields. For example, the train rail transit system established based on the plane multi-particle spring model \cite{ref1,ref2,ref3,ref4,ref5,ref6} and many simulations established by using the multi-particle spring branch model \cite{ref7,ref8,ref9} and so on. The most classic way to analyze spring multi-particle models is through force analysis combined with Newton's laws of kinematics. \cite{ref1,ref2,ref3,ref4,ref5,ref6,ref9}. This is also the way most people can understand the solution. However, if the number of particle systems is too large, and the connection between the particles is too complicated. Although classical Newtonian mechanics can still solve such problems, it is usually accompanied by problems,such as complicated derivation, large amounts of calculation, and error-prone. At this point, the use of analytical mechanics to solve the spring multi-particle model seems to be a good method.

	Take the compound pendulum system as an example \cite{ref10,ref11,ref12}. If the system is solved by classical Newtonian mechanics, we not only need to perform force analysis on two particle successively, but also need to perform a series of skillful methods such as reference frame transformation. This method is undoubtedly more cumbersome. The introduction of analytical mechanics can solve these problems ingeniously. It avoids the complicated force analysis, and only needs to construct the Lagrangian quantity of the system under suitable generalized coordinates and bring it into the Euler-Lagrangian equation to obtain the motion equation of the system. Although this method is simple in process, as the number of particles increases, the degree of freedom of the system increases. In this way, the Lagrangian quantity of the system will also become very complicated. At this time, even analytical mechanics will face a huge amount of computation.

	To solve this problem, Ref.\cite{ref9} proposes a method of constraint processing. That is one particle in the multi-branch spring particle system is taken as the research object, and other particle directly connected to it are taken as constraints. Then, write the relationship between the particle and the constraint term according to Newton's laws of motion. In the end, it is only necessary to do similar processing for each particle, and the constraints can be mapped to the entire particle system. This linear iterative processing method is also easy to program and process, and it is also more convenient to obtain the numerical results of the system. However, the problem of this method is that the force analysis is too complicated when the number of particles is large.
	
	Therefore, it is the key to solve this kind of problems whether to provide a simple and clear spring multi-particle dynamic model that is easy to deal with a large number of particles. In this work, we will study the dynamic model of multi-particle spring system under the framework of analytical mechanics.
	
	This paper is mainly divided into the following parts: First, we discuss the general form of the dynamic equations of the four-particle spring system for the plane and vertical cases in Sec. \ref{sec:A} and Sec. \ref{sec:B}, respectively. The physical quantity spring action $\mathcal{K}$ is defined from the derivation process, and the dynamic model of the multi-particle spring system is established based on $\mathcal{K}$. Later, in Sec. \ref{sec:C}, the dynamic model is extended to complex multi-particle spring systems such as closed-loop connections by introducing one-dimensional projection. Appendix.  \ref{app:1} presents three specific examples of numerical solutions and applications of our dynamic model. Finally, a summary of the whole work is given in Sec. \ref{sec:D}.

	\section{Model}
	\subsection{Planar spring multi-particle system}
	\label{sec:A}
\begin{figure}[h]
		\centering
		\includegraphics[width=1.0\linewidth]{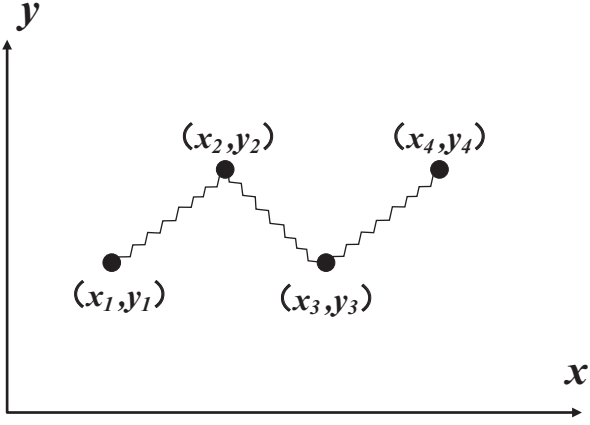}
		\caption{Planar four-particle spring system.}
		\label{fig:1}
	\end{figure}

	We start by deriving our model from the most basic planar four-particle spring system (Fig. \ref{fig:1}).
	
	First of all, some of the more important parameters of the system are agreed upon.Assume that the spring is a light spring with no mass, and the elastic stiffness is $k_j$, and the original length is $l_0^{(j)}$. The mass of the particle is $m_i$. And in a certain Cartesian coordinate system, each particle has an independent position coordinate $(x_i, y_i)$ at any time. In this way we can determine the length of each spring at any time $l_j(t)=\sqrt{(x_{n}-x_i)^2+(y_{n}-y_i)^2}$(Here $i$,$n$ and $j$ respectively represent the numbers of the particle and springs in the spring multi-particle system, which have no actual physical meaning).
	
	With these assumptions, the Lagrangian for a planar spring four-particle system shown in Fig. \ref{fig:1} is:
	\begin{equation}
		\begin{aligned}
			L&=T-V\\&=\frac{1}{2}m_1(\dot{x_1}^2+\dot{y_1}^2)+\frac{1}{2}m_2(\dot{x_2}^2+\dot{y_2}^2)+\frac{1}{2}m_3(\dot{x_3}^2+\dot{y_3}^2)\\&+\frac{1}{2}m_4(\dot{x_4}^2+\dot{y_4}^2)-\frac{1}{2}k_1(l_1-l_0^{(1)})^2-\frac{1}{2}k_2(l_2-l_0^{(2)})^2\\&-\frac{1}{2}k_3(l_3-l_0^{(3)})^2
			\label{equ:1}
		\end{aligned}
	\end{equation}
	In the case of no damping, the Lagrangian (\ref{equ:1}) satisfies $\frac{d}{dt} \frac{\partial L}{\partial \dot{q_\alpha}}-\frac{\partial L}{\partial q_\alpha}=0$ ,and the equations of motion of the plane spring four-particle system can be obtained by substituting the above formula and sorting it out:
	\begin{equation}
		\begin{cases}
			m_1\ddot{x_1}=k_1(x_2-x_1)(1-\frac{l_0^{(1)}}{l_1})\\
			m_1\ddot{y_1}=k_1(y_2-y_1)(1-\frac{l_0^{(1)}}{l_1})\\	m_2\ddot{x_2}=-k_1(x_2-x_1)(1-\frac{l_0^{(1)}}{l_1})+k_2(x_3-x_2)(1-\frac{l_0^{(2)}}{l_2})\\
			m_2\ddot{y_2}=-k_1(y_2-y_1)(1-\frac{l_0^{(1)}}{l_2})+k_2(y_3-y_2)(1-\frac{l_0^{(2)}}{l_2})\\
			m_3\ddot{x_3}=-k_2(x_3-x_2)(1-\frac{l_0^{(2)}}{l_2})+k_3(x_4-x_3)(1-\frac{l_0^{(3)}}{l_3})\\
			m_3\ddot{y_3}=-k_2(y_3-y_2)(1-\frac{l_0^{(2)}}{l_2})+k_3(y_4-y_3)(1-\frac{l_0^{(3)}}{l_3})\\
			m_4\ddot{x_4}=-k_3(y_4-y_3)(1-\frac{l_0^{(3)}}{l_3})\\
			m_4\ddot{y_4}=-k_3(y_4-y_3)(1-\frac{l_0^{(3)}}{l_3})
			\label{equ:2}
		\end{cases}
	\end{equation}

	It can be found from the calculation result (\ref{equ:2}) that the form of the equation system has strong symmetry. Each of these terms can be thought of as an algebraic sum of several products. To unify the form, we define this product form as the spring action. Its specific form is:
	\begin{equation}
		\begin{cases}
			\mathcal{K}_j^x=k_j(x_n-x_i)(1-\frac{l_0^{(j)}}{l_j})\\
			\mathcal{K}_j^y=k_j(y_n-y_i)(1-\frac{l_0^{(j)}}{l_j})
			\label{equ:3}
		\end{cases}
	\end{equation}

	Here we make a few notes about the defined spring action $\mathcal{K}$:

	1. As expressed by Eq. (\ref{equ:3}), the spring action $\mathcal{K}$ is a vector and satisfies the orthogonal decomposition. It reflects the action of the spring on a certain particle point connected in a certain direction. In other words, the spring action can be understood as a macroscopic representation of the spring's potential.

	2. We further expand  Eq. (\ref{equ:3}), taking the x component as an example:
	\begin{equation}
		\begin{aligned}
			\mathcal{K}_j^x&=k_j(x_n-x_i)(1-\frac{l_0^{(j)}}{l_j})\\&=k_j(x_n-x_i)(1-\frac{l_0^{(j)}}{\sqrt{(x_{n}-x_i)^2+(y_{n}-y_i)^2}})
			\label{equ:4}
		\end{aligned}	
	\end{equation}
	It can be found that the spring action $\mathcal{K}$ is a function of time $t$, which varies with the position of the particle connected to it. At a certain time, the magnitude of the spring action $\mathcal{K}$ is only determined by the elastic stiffness $k$ of the spring itself and the original length $l_0$, and has nothing to do with other factors. It can be seen that the amount of spring action can be regarded as an inherent property of the spring.

	Further research into Eq. (\ref{equ:2}) can be found. Although the equation of motion of the spring multi-particle system is the algebra and form of the action quantity, the signs connecting them are different. To solve this problem, we make the following definitions.

	According to the different tendency of the spring to make the particle close to or away from the coordinate origin in the extended state, we divide the way of the particle to connect the spring into active connection and passive connection.
	
	\begin{figure}[h]
		\centering
		\includegraphics[width=1.1\linewidth]{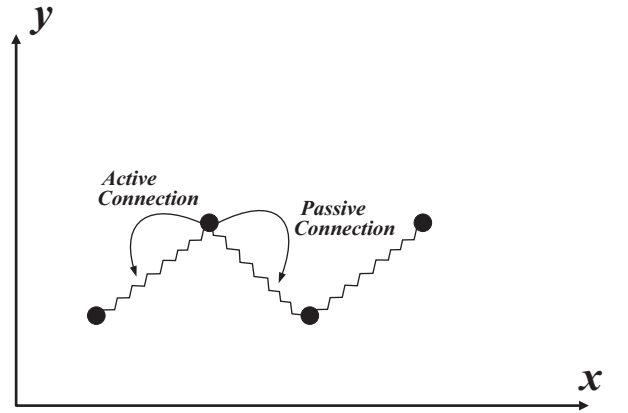}
		\caption{Schematic diagram of active connection and passive connection in the plane system.}
		\label{fig:2}
	\end{figure}

	Fig. \ref{fig:2} shows the schematic diagram of active connection and passive connection under the plane particle system. Take the second particle in the figure as an example: the spring connected on the left has a tendency to act on the particle close to the origin of coordinates in an extended state, which is called an active connection; the spring connected on the right has a tendency to move away from the origin of coordinates to the particle in an extended state. The role trend is called passive connection.

	For a certain particle, we define that the spring action of its active connection is negative(-$\mathcal{K}_{active}$), and the spring action of its passive connection is positive($\mathcal{K}_{passive}$). Taking the second particle in Fig.1 as an example, we rewrite its differential equation of motion in the form of spring action:
	\begin{equation}
		\begin{cases}
			m_2\ddot{x_2}=-\mathcal{K}_1^x+\mathcal{K}_2^x\\
			m_2\ddot{y_2}=-\mathcal{K}_1^y+\mathcal{K}_2^y
			\label{equ:5}
		\end{cases}
	\end{equation}

	However,  if the system is simultaneously acted by other dissipative forces $F$ other than the spring force, the form of the equations will not be significantly different. Because when there is a dissipative force in the system, the Euler-Lagrange equation it satisfies becomes$\frac{d}{dt}\frac{\partial L}{\partial \dot{q_\alpha}}-\frac{\partial L}{\partial q_\alpha}+\frac{\partial \mathcal{F}}{\partial \dot{q_\alpha}}=0$,where $\mathcal{F}$ is the dissipation function corresponding to the dissipation force. And the whole equation system is only the particle derivative of the dissipation function to the generalized velocity coordinate in form (The derivation of the dissipation function is not the focus of this article, and will not be explained here).

	So far, we have summarized the basic forms of the sequentially connected planar spring multi-particle model through the spring action:
	\begin{equation}
		\begin{cases}
			m_i\ddot{x_i}=\mathcal{K}_{passive}^x-\mathcal{K}_{active}^x+\frac{\partial \mathcal{F}}{\partial \dot{x_i}}\\
			m_i\ddot{y_i}=\mathcal{K}_{passive}^y-\mathcal{K}_{active}^y+\frac{\partial \mathcal{F}}{\partial \dot{y_i}}\\ \cdots
		\end{cases}
		\label{equ:6}
	\end{equation}

	This method is still applicable in the vertically placed spring pendulum particle system. Next, we will further verify the universality of this method.
	
	\subsection{Vertical spring pendulum multi-particle system}	
	\label{sec:B}

	Previously, we studied the spring multi-particle system in the plane coordinate system by the spring action. Next, we extend this method to a vertical spring pendulum multi-particle system.
	
	\begin{figure}[h]
		\centering
		\includegraphics[width=1.0\linewidth]{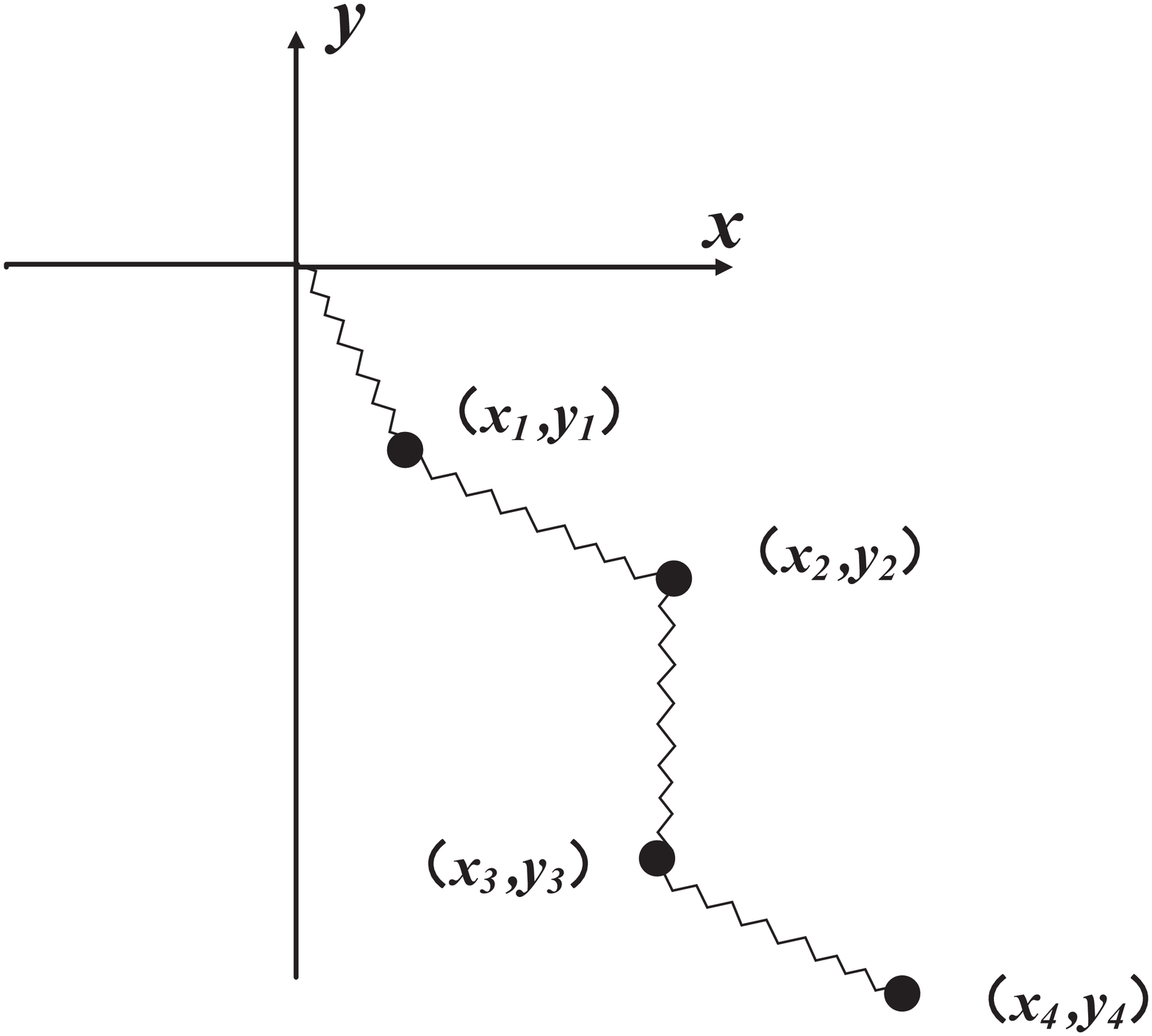}
		\caption{Four-particle spring pendulum.}
		\label{fig:3}
	\end{figure}
	
	We take the sequentially connected four-particle spring pendulum as an example (Fig. \ref{fig:3}), where the relevant parameter settings are the same as in Sec. \ref{sec:A}. First of all, when the whole system is undamped, we can easily write the Lagrangian of the system\cite{ref10,ref11,ref12}:
	\begin{equation}
		\begin{aligned}
			L&=T-V\\&=\frac{1}{2}m_1(\dot{x_1}^2+\dot{y_1}^2)+\frac{1}{2}m_2(\dot{x_2}^2+\dot{y_2}^2)+\frac{1}{2}m_3(\dot{x_3}^2+\dot{y_3}^2)\\&+\frac{1}{2}m_4(\dot{x_4}^2+\dot{y_4}^2)-\frac{1}{2}k_1(l_1-l_0^{(1)})^2-\frac{1}{2}k_2(l_2-l_0^{(2)})^2\\&-\frac{1}{2}k_3(l_3-l_0^{(3)})^2-\frac{1}{2}k_4(l_4-l_0^{(4)})^2-m_1gy_1-m_2gy_2\\&-m_3gy_3-m_4gy_4
			\label{equ:7}
		\end{aligned}
	\end{equation}
	and bring it into equation $\frac{d}{dt}\frac{\partial L}{\partial \dot{q_\alpha}}-\frac{\partial L}{\partial q_\alpha}=0$ to get the equations of motion of the four-particle spring pendulum:
	\begin{equation}
		\begin{cases}
			m_1\ddot{x_1}=-k_1x_1(1-\frac{l_0^{(1)}}{l_1})+k_2(x_2-x_1)(1-\frac{l_0^{(2)}}{l_2})\\
			
			m_1\ddot{y_1}=-k_1y_1(1-\frac{l_0^{(1)}}{l_1})+k_2(y_2-y_1)(1-\frac{l_0^{(2)}}{l_2})-m_1g\\
			
			m_2\ddot{x_2}=-k_2(x_2-x_1)(1-\frac{l_0^{(2)}}{l_2})+k_3(x_3-x_2)(1-\frac{l_0^{(3)}}{l_3})\\
			
			m_2\ddot{y_2}=-k_2(y_2-y_1)(1-\frac{l_0^{(2)}}{l_2})+k_3(y_3-y_2)(1-\frac{l_0^{(3)}}{l_3})-m_2g\\
			
			m_3\ddot{x_3}=-k_3(y_3-y_2)(1-\frac{l_0^{(3)}}{l_3})+k_4(x_4-x_3)(1-\frac{l_0^{(4)}}{l_4})\\
			
			m_3\ddot{y_3}=-k_3(y_3-y_2)(1-\frac{l_0^{(3)}}{l_3})+k_4(x_4-x_3)(1-\frac{l_0^{(4)}}{l_4})-m_3g\\
			
			m_4\ddot{x_4}=-k_4(x_4-x_3)(1-\frac{l_0^{(4)}}{l_4})\\
			
			m_4\ddot{y_4}=-k_4(y_4-y_3)(1-\frac{l_0^{(4)}}{l_4})-m_4g
			
			\label{equ:8}
		\end{cases}
	\end{equation}

	Compared with Eq. (\ref{equ:2}), it is found that there is no difference in form between the two equations. The difference is that the spring pendulum system has an additional gravity term $m_ig$ on the $y$ component.
	
\begin{figure}[h]
		\centering
		\includegraphics[width=0.8\linewidth]{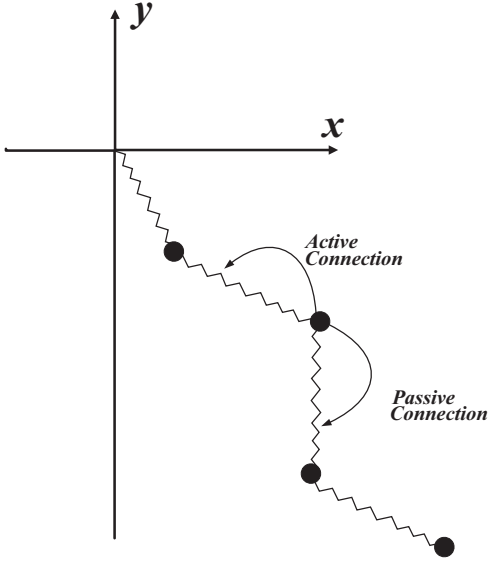}
		\caption{Schematic diagram of active connection and passive connection of vertical multi-particle spring pendulum system.}
		\label{fig:4}
	\end{figure}

	Therefore, for the spring pendulum multi-particle system, we can also use the spring action quantity Eq. (\ref{equ:4}) to express. And using the concept of active connection and passive connection mentioned earlier, the sign problem between the actions can also be determined (Fig. \ref{fig:4}).
	
	It should be noted that for the first particle in the spring pendulum multi-particle system (Fig. \ref{fig:3}), the other side of the spring that is actively connected to it is connected to the connection point on the wall. At this time, we can't assume that this connection point does not belong to the multi-particle system, but we need to regard the connection point as a fixed point with coordinates $(0,0)$ in the particle system. So the action of the first spring expands as(using the $x$ component as an example):
	\begin{equation}
		\begin{aligned}
			\mathcal{K}_1^x&=k_1(x_1-0)(1-\frac{l_0^{(1)}}{\sqrt{(x_{1}-0)^2+(y_1-0)^2}})\\&=k_1x_1(1-\frac{l_0^{(1)}}{\sqrt{x_{1}^2+y_1^2}})
		\end{aligned}
	\end{equation}

\begin{figure}[h]
		\centering
		\includegraphics[width=1.0\linewidth]{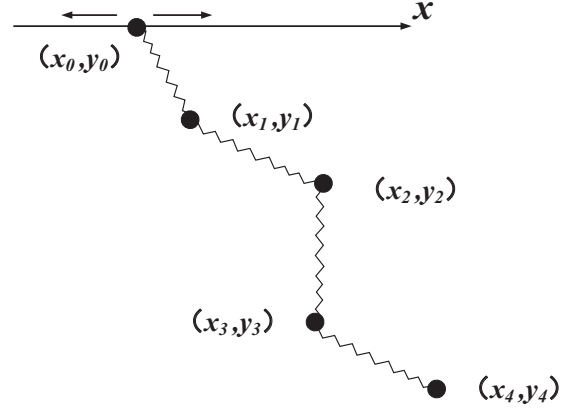}
		\caption{Spring pendulum with loose connection point.}
		\label{fig:5}
	\end{figure}

	This explanation is particularly important in a spring pendulum multi-particle system where the connection points are not fixed (Fig. \ref{fig:5}). Otherwise, ignoring the connection points will usually make the whole equation wrong.
	
	For the spring pendulum with damping, the treatment method is exactly the same as in Sec. \ref{sec:A}. That is, the equation system has an additional particle derivative of the dissipation function to the generalized velocity. No further elaboration here.

	So we can get the action representation of the spring pendulum multi-particle system:
	\begin{equation}
		\begin{cases}
			m_i\ddot{x_i}=\mathcal{K}_{passive}^x-\mathcal{K}_{active}^x+\frac{\partial \mathcal{F}}{\partial \dot{x_i}}\\
			m_i\ddot{y_i}=\mathcal{K}_{passive}^y-\mathcal{K}_{active}^y+\frac{\partial \mathcal{F}}{\partial \dot{y_i}}-m_ig\\ \cdots
			\label{equ:10}
		\end{cases}
	\end{equation}
	\subsection{complex connection system.}
	\label{sec:C}
	Through the previous derivation, it is found that the spring action we define can greatly simplify the calculation when describing the sequentially connected system. Now we introduce the spring action into the spring multi-particle system with complex connection (such as closed-loop connection, mesh connection, etc.) and discuss its convenience for complex system calculation.

\begin{figure}[h]
		\centering
		\includegraphics[width=1.0\linewidth]{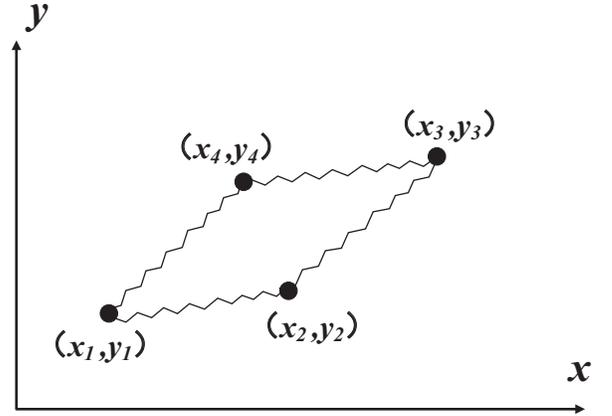}
		\caption{Closed-loop connected multi-particle spring system.}
		\label{fig:6}
	\end{figure}

	We take a closed-loop connected planar spring particle system as an example (Fig. \ref{fig:6}). If it is difficult to directly derive its motion equations, however, we can directly write its motion equations by defining the spring action. Specific steps are as follows.
	
	For an arbitrarily connected complex system. We firstly project the particles in the particle system onto a straight line by one-dimensional projection, so as to judge whether the particles satisfy the active connection or the passive connection. Furthermore, the motion differential equations of the particle group are written out through the action relationship between the springs.
	
	\begin{figure*}[htbp]
		\centering
		\includegraphics[width=1.0\linewidth]{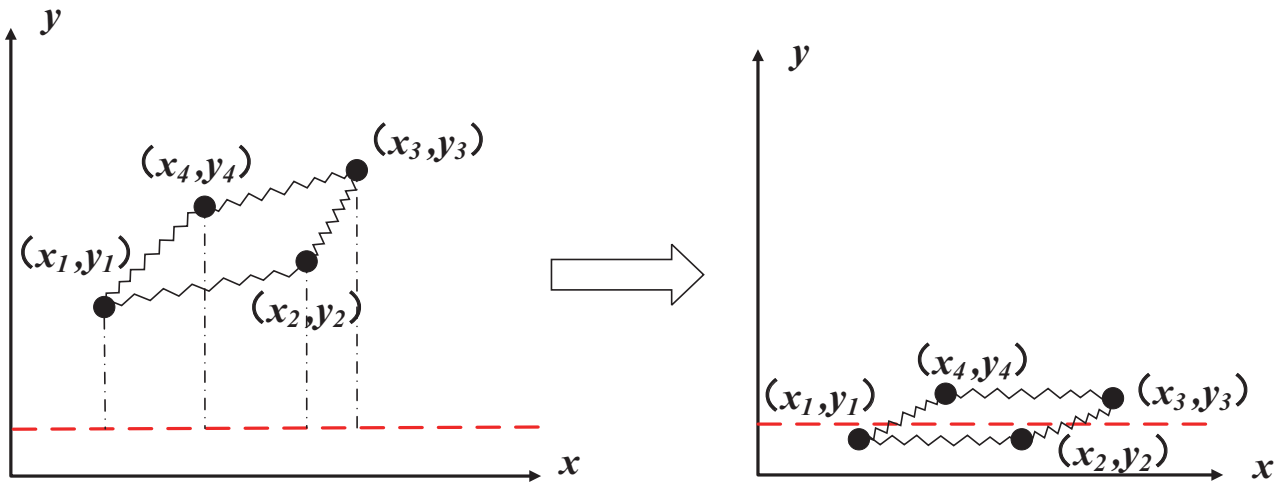}
		\caption{One-dimensional projection of complex multi-particle spring system.}
		\label{fig:7}
	\end{figure*}

\begin{figure*}[htbp]
		\centering
		\includegraphics[width=1.0\linewidth]{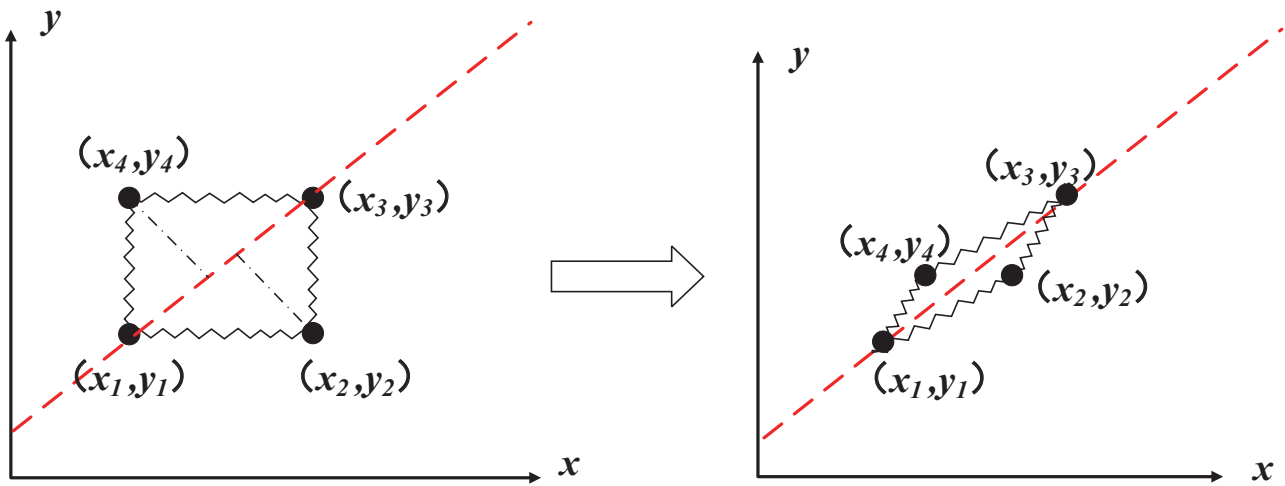}
		\caption{The direction of the one-dimensional projection of the particle system is always along the direction that is convenient to distinguish the connection between the particle and the spring}
		\label{fig:8}
	\end{figure*}

	The schematic diagram of one-dimensional projection is shown in Fig. \ref{fig:7}. We project a closed-loop system along a certain direction. It should be noted that this projection direction is not unique. According to the different initial positions of the complex particle system,.The one-dimensional projection direction of the particle system is always along the direction that is convenient for distinguishing the connection between the particle and the spring (Fig. \ref{fig:8}). The reason for this is that, without changing the connection order of the springs between the particles, we can easily determine the connection between the particles and the springs by means of projection. After that, set the relative initial coordinates for the particle points, and then the initial state of the particle system can be restored.

	On this basis, we can determine the connection between the particle and the spring. For example, the No.1 particle in Fig. \ref{fig:7} is passively connected to the No.4 particle; the No.2 particle is actively connected to the No.1 particle, and is passively connected to the No.3 particle. And so on. In this way, by listing the relationship between the spring actions, we can directly write the differential equations of motion for the closed-loop system:
	\begin{equation}
		\begin{cases}
			m_1\ddot{x_1}=k_1(x_2-x_1)(1-\frac{l_0^{(1)}}{l_1})+k_4(x_4-x_1)(1-\frac{l_0^{(1)}}{l_4})\\
			m_1\ddot{y_1}=k_1(y_2-y_1)(1-\frac{l_0^{(1)}}{l_1})+k_4(y_4-y_1)(1-\frac{l_0^{(1)}}{l_4})\\	m_2\ddot{x_2}=-k_1(x_2-x_1)(1-\frac{l_0^{(1)}}{l_1})+k_2(x_3-x_2)(1-\frac{l_0^{(2)}}{l_2})\\
			m_2\ddot{y_2}=-k_1(y_2-y_1)(1-\frac{l_0^{(1)}}{l_2})+k_2(y_3-y_2)(1-\frac{l_0^{(2)}}{l_2})\\
			m_3\ddot{x_3}=-k_2(x_3-x_2)(1-\frac{l_0^{(2)}}{l_2})-k_3(x_3-x_4)(1-\frac{l_0^{(3)}}{l_3})\\
			m_3\ddot{y_3}=-k_2(y_3-y_2)(1-\frac{l_0^{(2)}}{l_2})-k_3(y_3-y_4)(1-\frac{l_0^{(3)}}{l_3})\\
			m_4\ddot{x_4}=-k_4(x_4-x_1)(1-\frac{l_0^{(1)}}{l_4})+k_3(x_3-x_4)(1-\frac{l_0^{(3)}}{l_3})\\
			m_4\ddot{y_4}=-k_4(y_4-y_1)(1-\frac{l_0^{(1)}}{l_4})+k_3(y_3-y_4)(1-\frac{l_0^{(3)}}{l_3})
			\label{equ:11}
		\end{cases}
	\end{equation}

	At this point, we can push the spring multi-particle model to the general case:
	\begin{equation}
		\begin{cases}
			m_i\ddot{x_i}=\sum\mathcal{K}_{passive}^x-\sum\mathcal{K}_{active}^x+\frac{\partial \mathcal{F}}{\partial \dot{x_i}}\\
			m_i\ddot{y_i}=\sum\mathcal{K}_{passive}^y-\sum\mathcal{K}_{active}^y+\frac{\partial \mathcal{F}}{\partial \dot{y_i}}\\ \cdots
			\label{equ:12}
		\end{cases}
	\end{equation}

	Visible, for any system connected by springs. We can all directly write the differential equations of motion of the system by defining the amount of action of the spring on the particles and how they are connected. This approach greatly simplifies the derivation of equations for complex systems and multi-particle spring systems. At the same time, the programming idea of this method is also very simple. It only needs to set the relevant basic parameters in the particle system, and use the calculation software to directly solve the relevant laws in the particle system. Therefore, our method can simplify the original spring model when solving many engineering problems with similar multi-particle models as the theoretical basis, so as to achieve the effect of getting twice the result with half the effort.Regarding the specific calculation process, we will not describe too much here. The specific example calculation of the model is shown in Appendix \ref{app:1}.

 \section{summary}
	\label{sec:D}
	In this work,we define a new physical quantity from the analysis of the simple spring multi-particle model $\mathcal{K}$.
	
	$\mathcal{K}$ is a mechanical vector whose direction is always along the direction of the spring, and its magnitude varies with time as the position coordinates of the connected particles. And it is found that for any multi-particle spring system, its equation form is actually represented by the algebraic sum of the spring action. What is difficult to determine, however, is the sign between the actions. We solve this problem neatly by defining the connection between the spring and the particle.
	
	We define that for a spring in an elongated state, if the spring has a tendency to make the mass connected to it move closer to the origin, then we call the mass connected to the spring an active connection; conversely, if the spring has a tendency to make the mass connected to it move away from the origin, then we call the mass connected to the spring a passive connection. And it is stipulated that the spring action amount corresponding to the active connection takes a negative sign, and the spring action amount of the passive connection takes a positive sign.
	
	For the sequential connection systems shown in Sec. \ref{sec:A} and Sec. \ref{sec:B}. The active and passive connection between the spring and the particle is relatively easy to determine. But for complex connection systems (such as closed-loop connections, etc.), active and passive connections are not easy to determine. At this time, we perform one-dimensional projection of the complex system: on the premise that the connection relationship between the particle and the spring remains unchanged, the particle is projected onto a one-dimensional space that is easy to distinguish between active and passive connections (Fig. \ref{fig:7}). It should be noted that the direction of the one-dimensional space depends on the initial arrangement of the complex system, and it is always along the direction that facilitates the distinction between the connections(Fig.  \ref{fig:8}).

	In this way, the motion equations (Eq. (\ref{equ:12} )) of the spring particle system can be expressed by the spring action. So far, a new dynamic model of the multi-particle spring system has been established, which is not only makes the derivation of the equations of motion of the spring system extremely simple, but also easier to program. This makes the calculation of numerical solutions of spring multi-particle models extremely simple compared to traditional methods. Especially when calculating a system with a large number of particles, it can better reflect the superiority of this system.
	
	\section{Acknowledgments}

This work is supported by the National Natural Science Foundation of China under Grant Nos. 12065014 and 12047501, and by the Natural Science Foundation of Gansu province under Grant No. 22JR5RA266. We acknowledge the West Light Foundation of The Chinese Academy of Sciences, Grant No. 21JR7RA201.

	\clearpage
	
	\appendix
	\section{Numerical calculation example}
	\label{app:1}
	\subsection{Ten-particle spring pendulum}
	The solution of traditional spring multi-particle model is often accompanied by difficult equation derivation and complicated program design. And our proposed spring action model overcomes these two difficulties very well.

	\begin{figure}[h]
		\centering
		\includegraphics[width=1.0\linewidth]{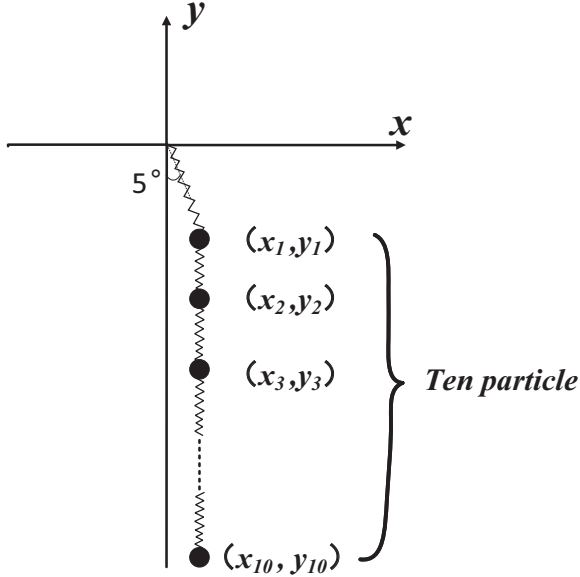}
		\caption{Ten-particle spring pendulum system.}
		\label{fig:9}
	\end{figure}
	
	For the first example, we choose a spring pendulum model consisting of ten particles (Fig. \ref{fig:9}). For the sequentially connected connected spring pendulum model, we can directly determine the connection relationship between each pair of springs and particles Sec. \ref{sec:B}. Then list the equations of motion of the system according to Eq. (\ref{equ:10})
	\begin{equation}
		\begin{cases}
			m_1\ddot{x_1}=-k_1x_1(1-\frac{l_0^{(1)}}{l_1})+k_2(x_2-x_1)(1-\frac{l_0^{(2)}}{l_2})\\
			
			m_1\ddot{y_1}=-k_1y_1(1-\frac{l_0^{(1)}}{l_1})+k_2(y_2-y_1)(1-\frac{l_0^{(2)}}{l_2})-m_1g\\
			
			m_2\ddot{x_2}=-k_2(x_2-x_1)(1-\frac{l_0^{(2)}}{l_2})+k_3(x_3-x_2)(1-\frac{l_0^{(3)}}{l_3})\\
			
			m_2\ddot{y_2}=-k_2(y_2-y_1)(1-\frac{l_0^{(2)}}{l_2})+k_3(y_3-y_2)(1-\frac{l_0^{(3)}}{l_3})-m_2g\\
			
			\cdots\\
			m_9\ddot{x_9}=-k_9(x_9-x_8)(1-\frac{l_0^{(9)}}{l_9})+k_{10}(x_{10}-x_9)(1-\frac{l_0^{(10)}}{l_{10}})\\
			
			m_9\ddot{y_9}=-k_9(y_9-y_8)(1-\frac{l_0^{(9)}}{l_9})+k_{10}(y_{10}-y_9)(1-\frac{l_0^{(10)}}{l_{10}})-m_9g\\
			m_{10}\ddot{x_{10}}=-k_{10}(x_{10}-x_9)(1-\frac{l_0^{({10})}}{l_{10}})\\
			
			m_{10}\ddot{y_{10}}=-k_{10}(y_{10}-y_9)(1-\frac{l_0^{({10})}}{l_{10}})-m_{10}g
			
			\label{equ:13}
		\end{cases}
	\end{equation}
	where $l_i=\sqrt{(x_{i}-x_{i-1})^2+(y_{i}-y_{i-1})^2}$

\begin{figure}[htbp]
		\centering
		\begin{minipage}{0.48\linewidth}
			\centering
			\includegraphics[width=1.0\linewidth]{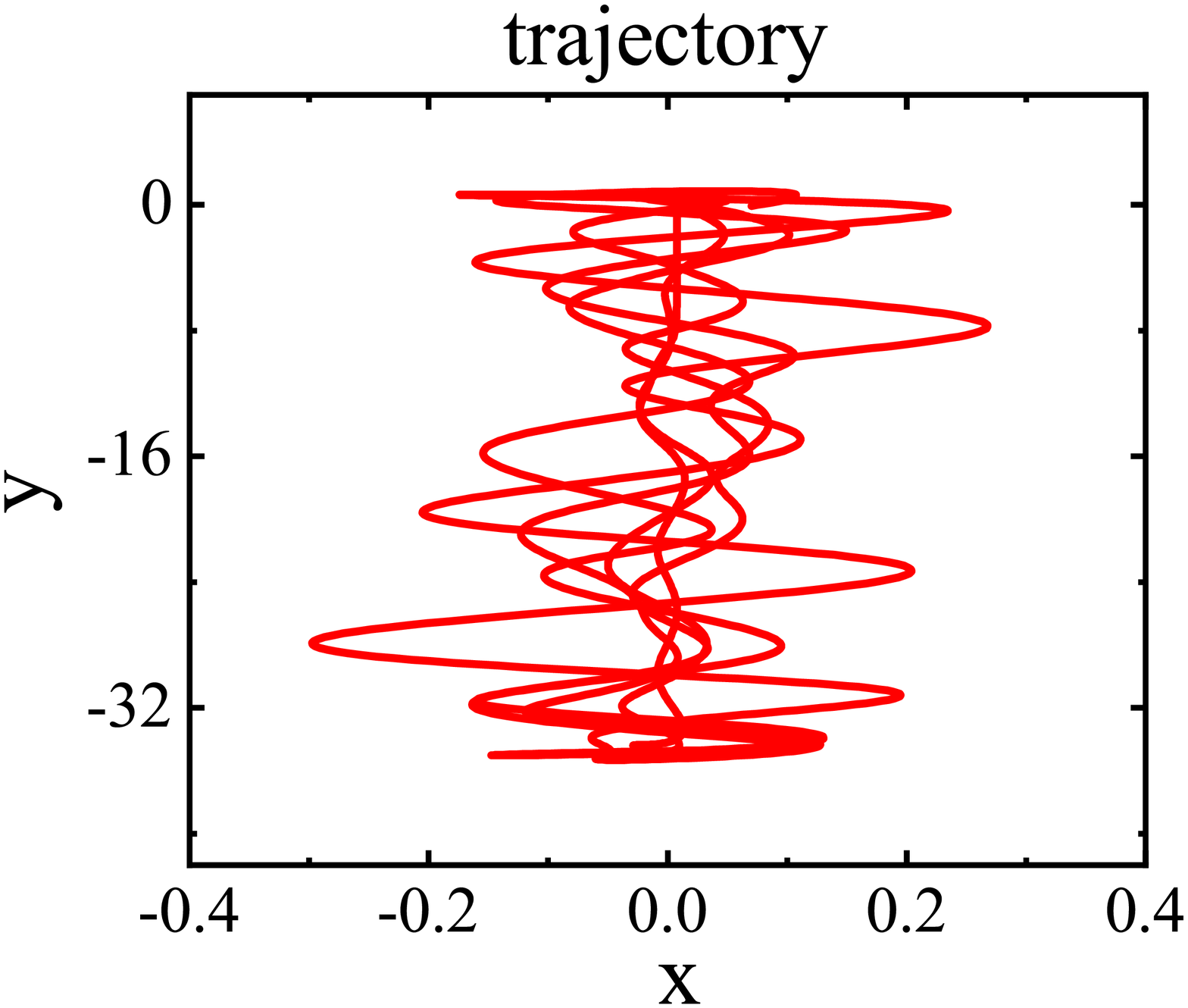}
		\end{minipage}
		\begin{minipage}{0.48\linewidth}
			\centering
			\includegraphics[width=1.0\linewidth ]{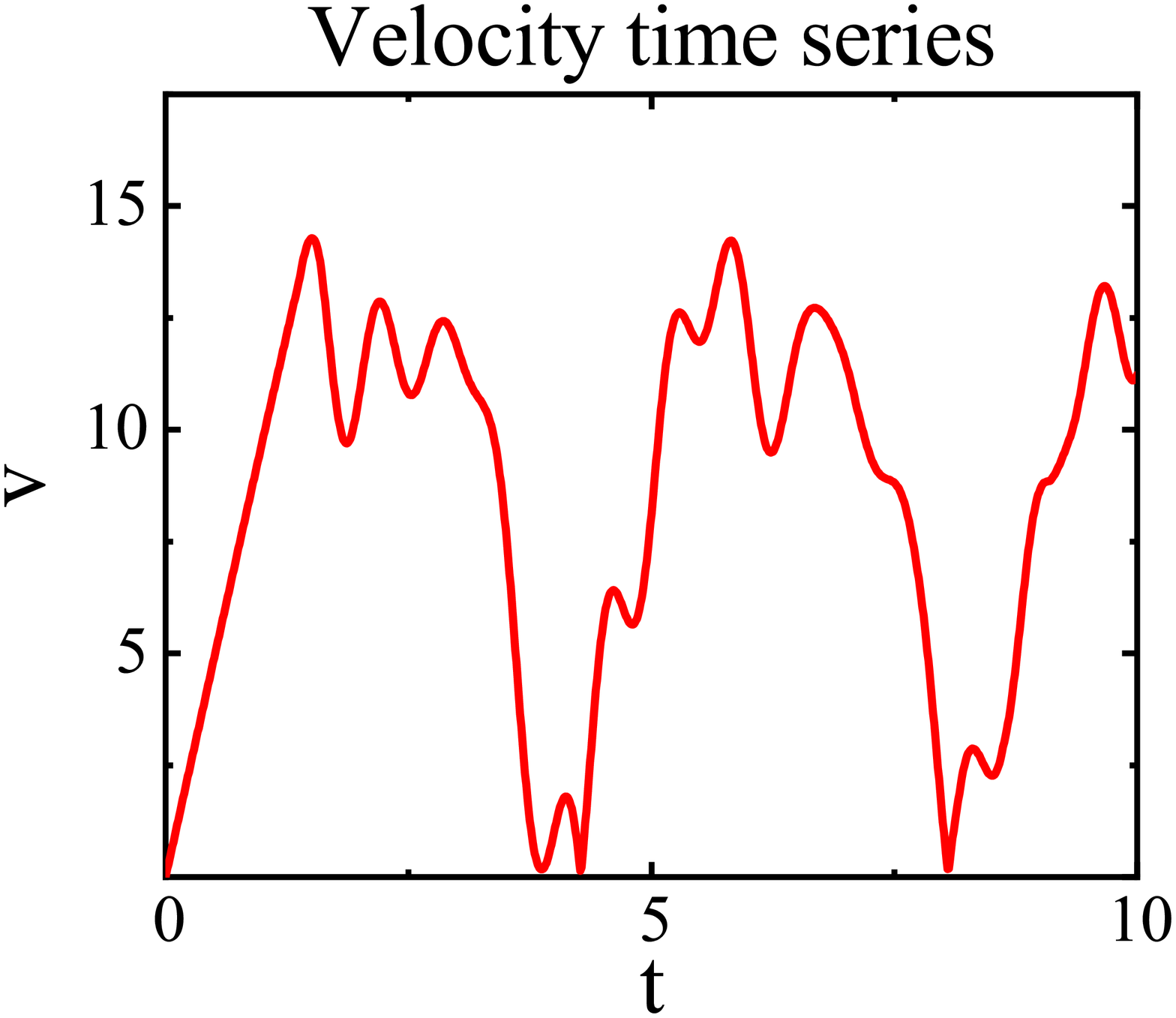}
		\end{minipage}
		\qquad
		\begin{minipage}{0.48\linewidth}
			\centering
			\includegraphics[width=1.0\linewidth]{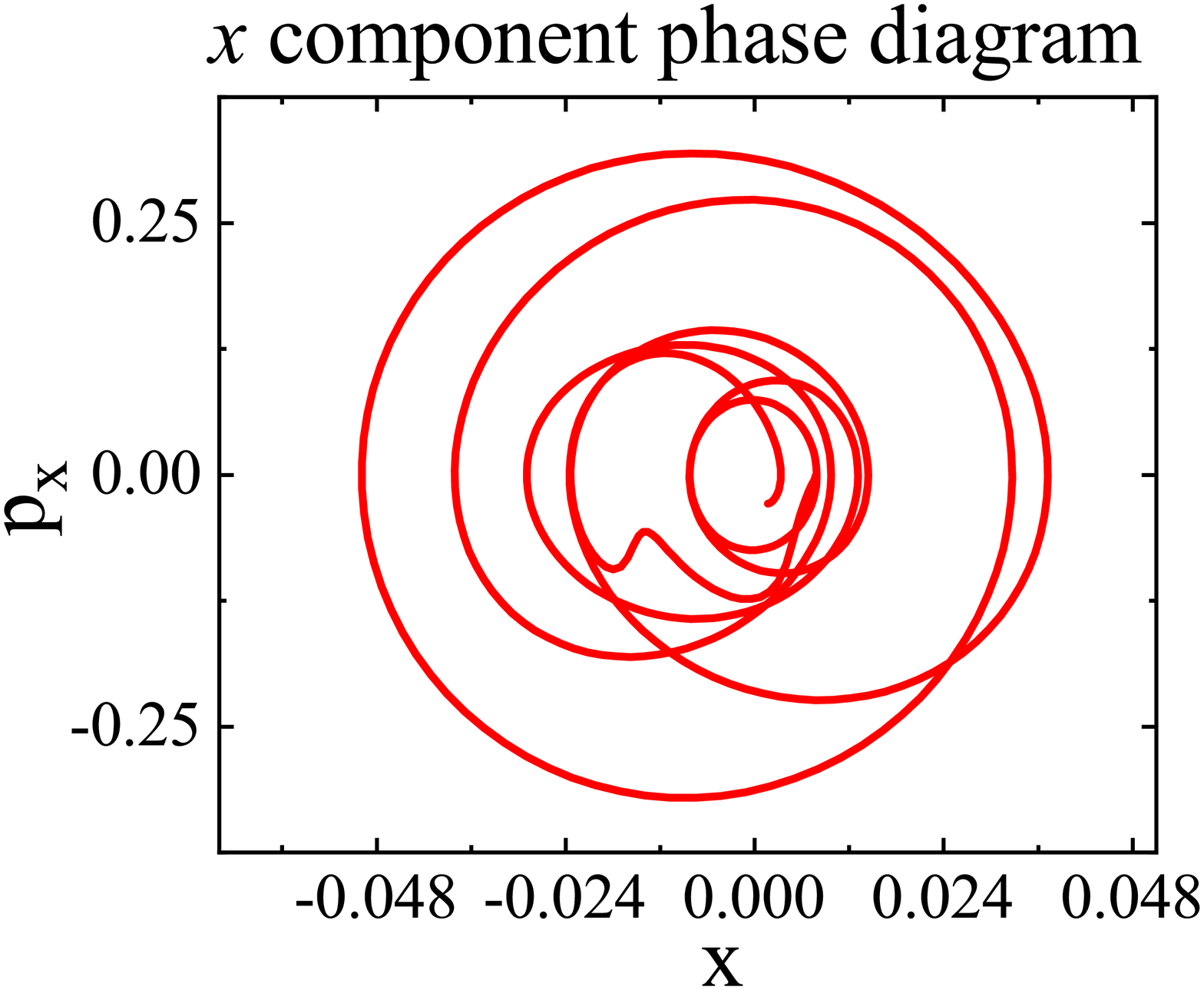}
		\end{minipage}
		\begin{minipage}{0.48\linewidth}
			\centering
			\includegraphics[width=1.0\linewidth]{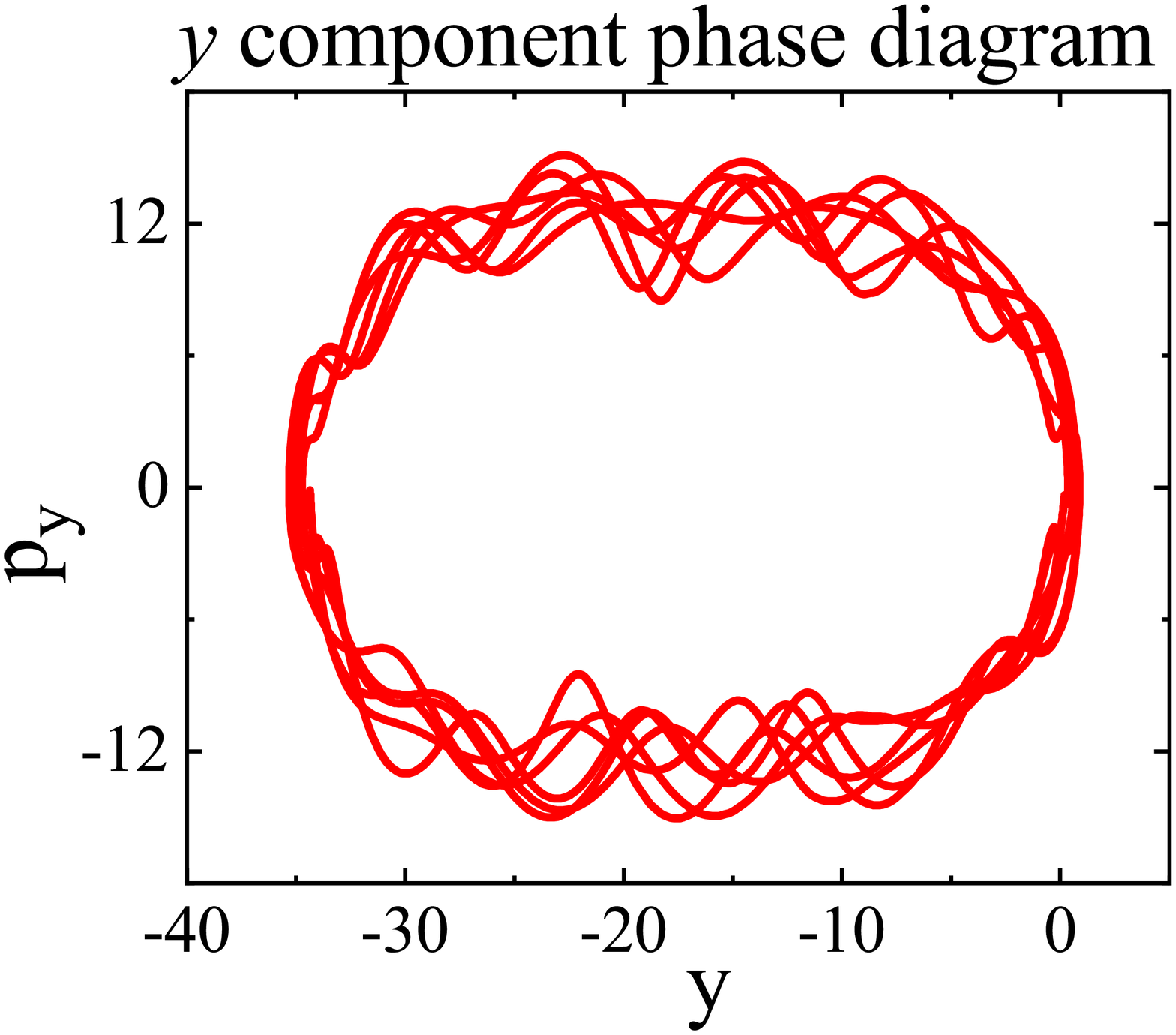}
		\end{minipage}
		\caption{{Example of the calculation result of the motion state of the sixth particles in the ten-particle spring pendulum.}}
		\label{fig:j1}
	\end{figure}

	The initial conditions for the spring ten pendulum are set. We assume that the angle between the first particle and the $y$-axis is $5^{\circ}$, and the rest of the particles are placed in the vertical direction (Fig.\ref{fig:9}). After that, through simple programming calculations, we can easily get the motion state of each particle in the system. Including motion trajectories, phase diagrams and time series .Here we show the numerical results of the sixth particle in the system as an example (Fig.\ref{fig:j1})
	
	\subsection{Planar ten-particle spring system}
\begin{figure}
		\centering
		\includegraphics[width=1.0\linewidth]{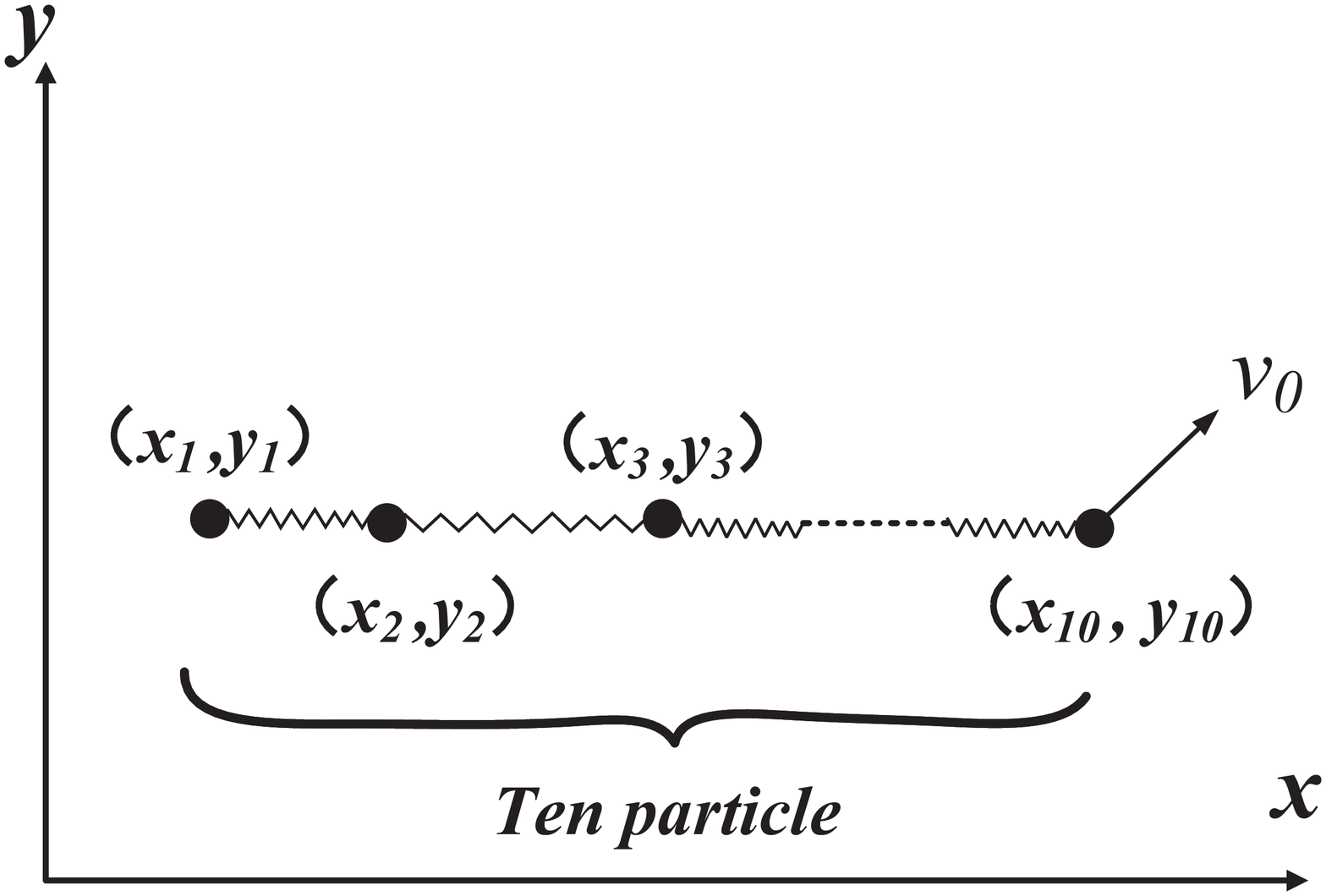}
		\caption{Panar ten-particle spring system.}
		\label{fig:10}
	\end{figure}

	Then we take the 10-particles spring system in the plane as an example to further verify the superiority of the spring action model. We let the initial arrangement of the particle system lie on a horizontal line, and each spring has a certain initial potential energy. At the same time, the tenth particle is given an initial velocity deviating from the horizontal line (Fig. \ref{fig:10}). To be more realistic, it is assumed that the system is subjected to linear damping related to the velocity of the particle itself, denoted as: $f=cv$.

	For the sequentially connected system mentioned in Sec. \ref{sec:A}. From left to right, we determine that the spring connected to the left of each particle is an active connection, and the spring connected to the right is a passive connection. In this way, according to Eq. (\ref{equ:4}) and Eq. (\ref{equ:6}), the motion differential equations of the plane spring particle system can be directly written:
	
	\begin{equation}
		\begin{cases}
			m_1\ddot{x_1}=k_1(x_2-x_1)(1-\frac{l_0^{(2)}}{l_2})-c\dot{x_1}\\
			
			m_1\ddot{y_1}=k_1(y_2-y_1)(1-\frac{l_0^{(2)}}{l_2})-c\dot{y_1}\\
			
			m_2\ddot{x_2}=-k_1(x_2-x_1)(1-\frac{l_0^{(2)}}{l_2})+k_2(x_3-x_2)(1-\frac{l_0^{(3)}}{l_3})-c\dot{x_2}\\
			
			m_2\ddot{y_2}=-k_1(x_2-x_1)(1-\frac{l_0^{(2)}}{l_2})+k_2(y_3-y_2)(1-\frac{l_0^{(3)}}{l_3})-c\dot{y_2}\\
			
			\cdots\\
			m_9\ddot{x_9}=-k_8(x_9-x_8)(1-\frac{l_0^{(8)}}{l_8})+k_{9}(x_{10}-x_9)(1-\frac{l_0^{(9)}}{l_{9}})-c\dot{x_9}\\
			
			m_9\ddot{y_9}=-k_8(y_9-y_8)(1-\frac{l_0^{(8)}}{l_8})+k_{9}(y_{10}-y_9)(1-\frac{l_0^{(9)}}{l_{9}}-c\dot{y_9}\\
			m_{10}\ddot{x_{10}}=-k_{9}(x_{10}-x_9)(1-\frac{l_0^{({9})}}{l_{9}})-c\dot{x_{10}}\\
			
			m_{10}\ddot{y_{10}}=-k_{9}(y_{10}-y_9)(1-\frac{l_0^{({9})}}{l_{9}})-c\dot{y_{10}}
			
			\label{equ:14}
		\end{cases}
	\end{equation}

	Similarly, the motion state of any particle in the particle system is calculated by software. Here is an example of the eighth particle point(Fig. \ref{fig:j2}):
	\begin{figure}[htbp]
		\centering
		\begin{minipage}{0.48\linewidth}
			\centering
			\includegraphics[width=1.0\linewidth]{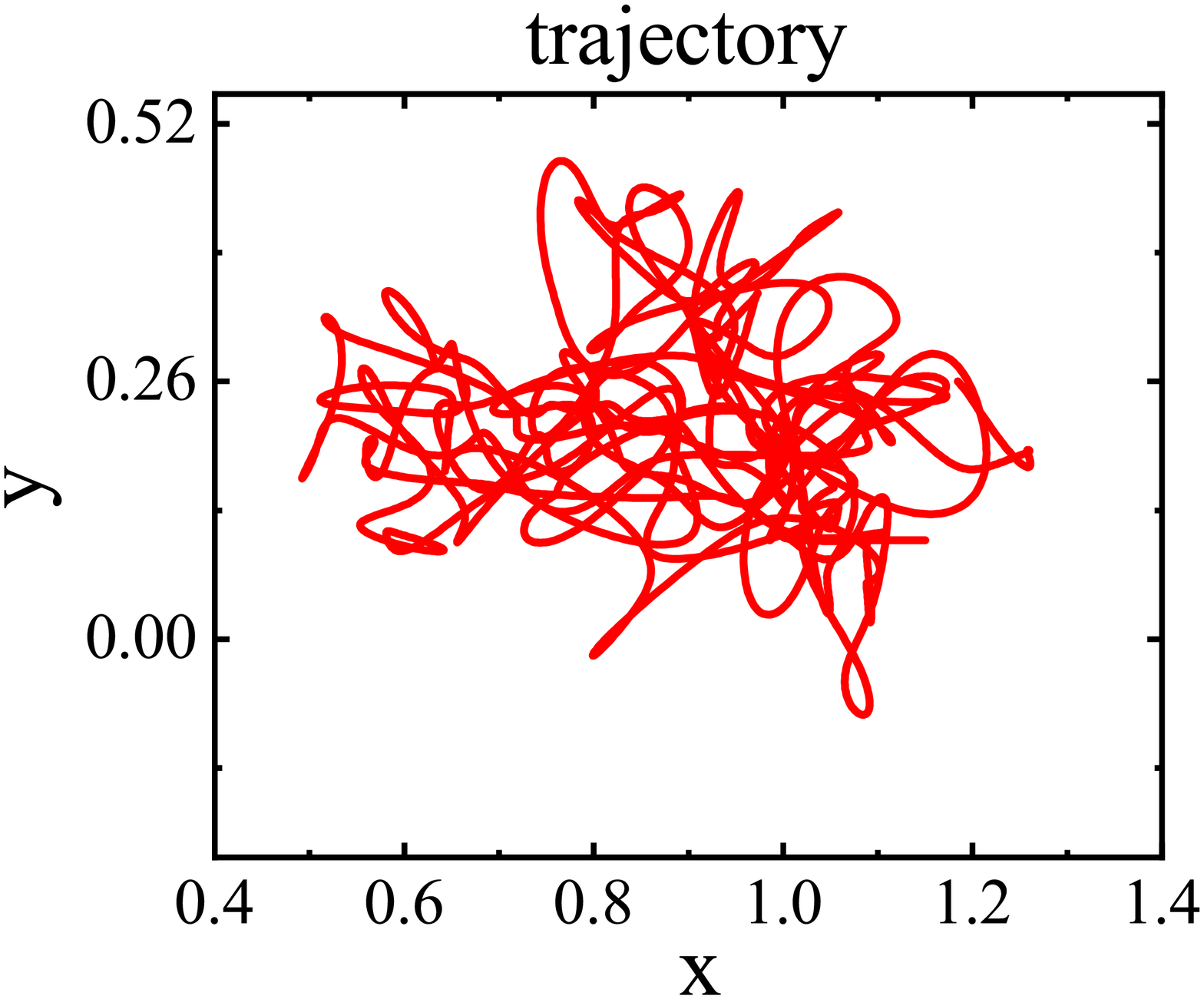}
		\end{minipage}
		\begin{minipage}{0.48\linewidth}
			\centering
			\includegraphics[width=1.0\linewidth ]{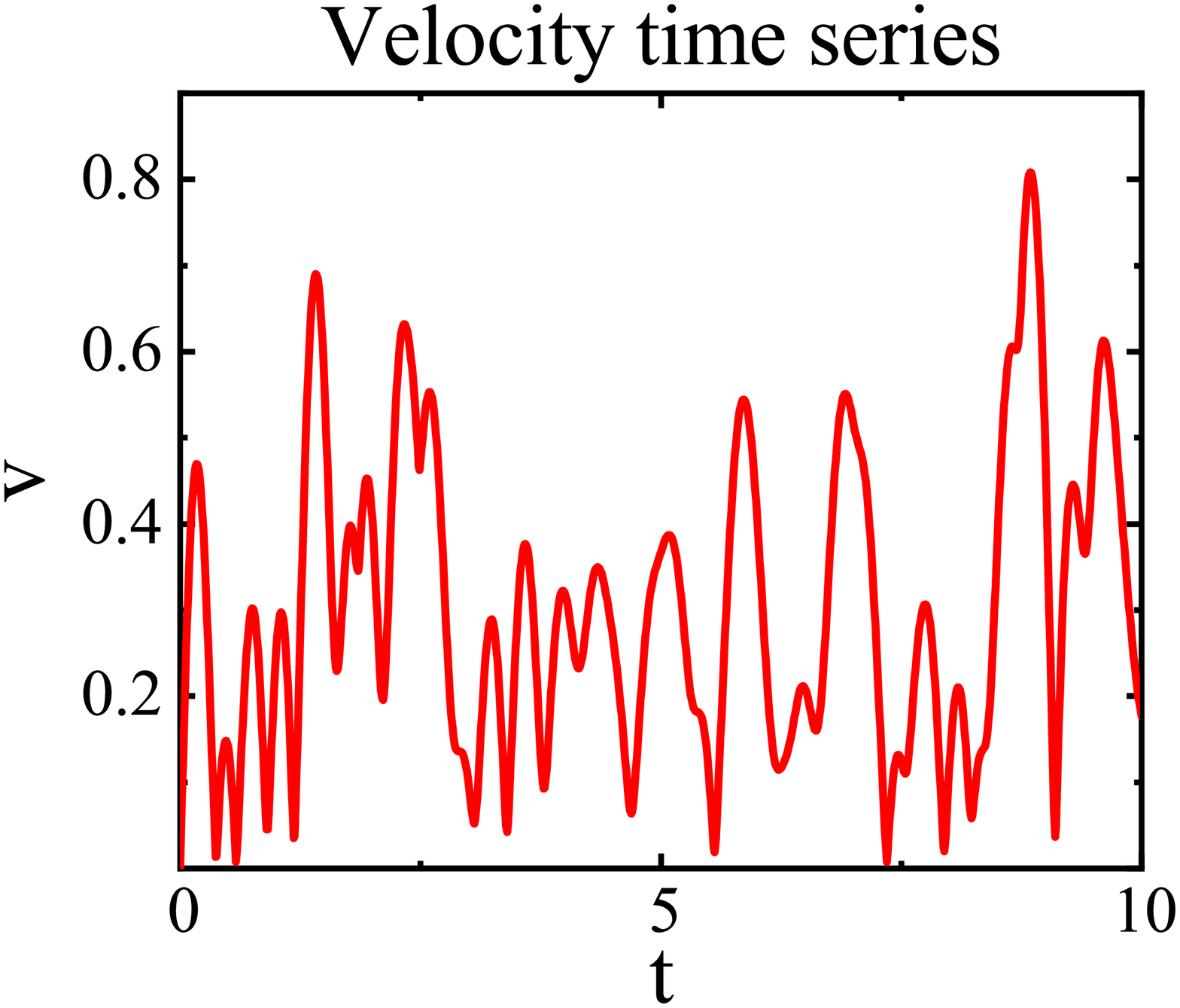}
		\end{minipage}
		\qquad
		\begin{minipage}{0.48\linewidth}
			\centering
			\includegraphics[width=1.0\linewidth]{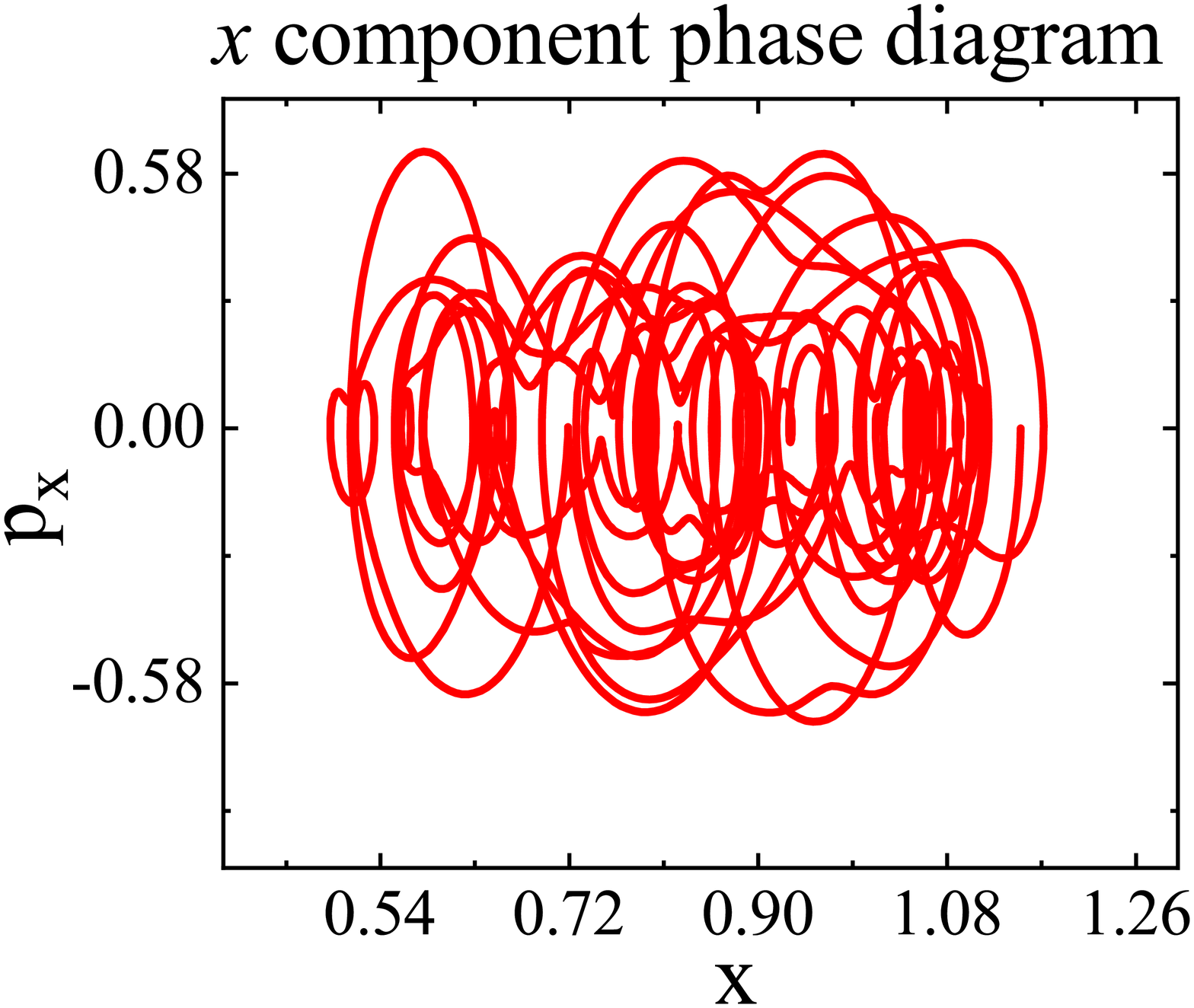}
		\end{minipage}
		\begin{minipage}{0.48\linewidth}
			\centering
			\includegraphics[width=1.0\linewidth]{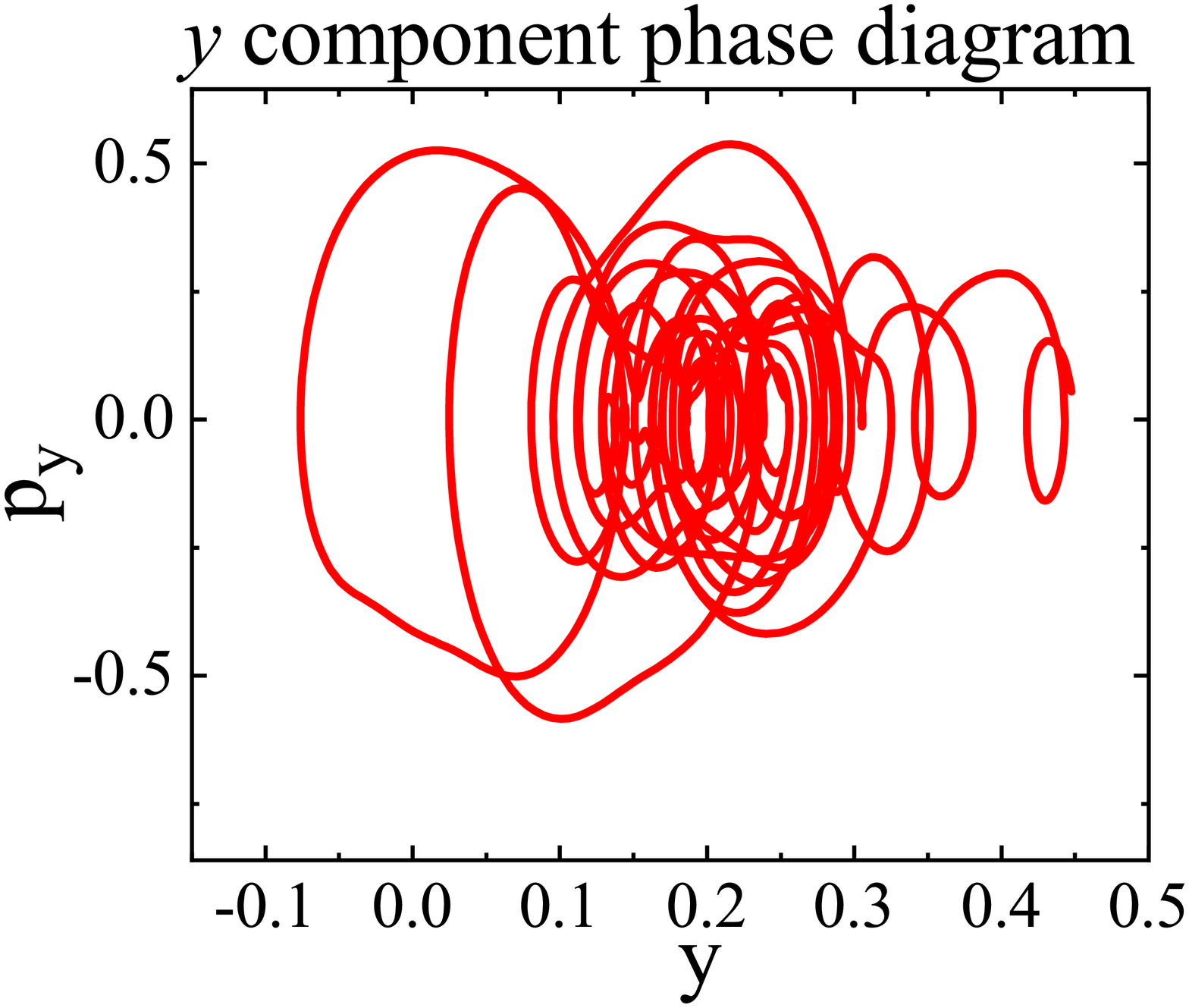}
		\end{minipage}
		\caption{{Example of the calculation result of the motion state of the sixth particles in the ten-particles spring pendulum.}}
		\label{fig:j2}
	\end{figure}
	
	\subsection{closed loop system}
	Finally, for complex connected systems. Compared with the previous two cases, the processing method is only one more one-dimensional projection process, and the other steps are the same. Here we take the closed-loop system shown in Sec. \ref{sec:C} as an example to further illustrate the calculation.

	As shown in Fig. \ref{fig:7}:

	\emph{step 1:} projects the particles in the plane onto a one-dimensional space parallel to the $x$-axis. The purpose of this is to determine whether the particle and the spring are connected actively or passively without changing the connection sequence between the particle and the spring. This is used to determine the sign of the action of the spring on the particle.

	\emph{Step 2:} According to Eq. (\ref{equ:4} )and Eq. (\ref{equ:12}), directly write the motion differential equations of the closed-loop system Eq. (\ref{equ:11}).

\begin{figure}[htbp]
		\centering
		\begin{minipage}{0.48\linewidth}
			\centering
			\includegraphics[width=1.0\linewidth]{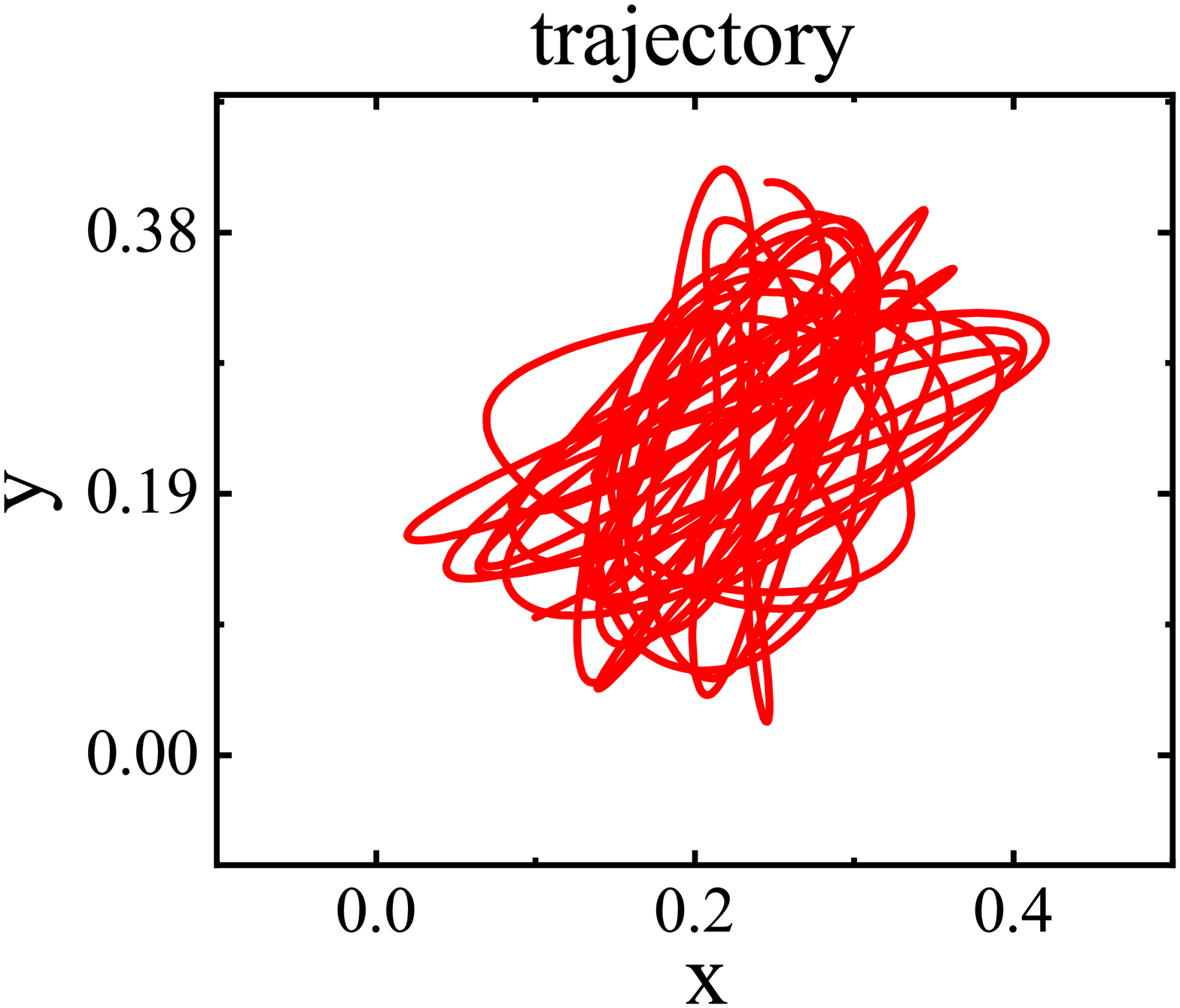}
		\end{minipage}
		\begin{minipage}{0.48\linewidth}
			\centering
			\includegraphics[width=1.0\linewidth ]{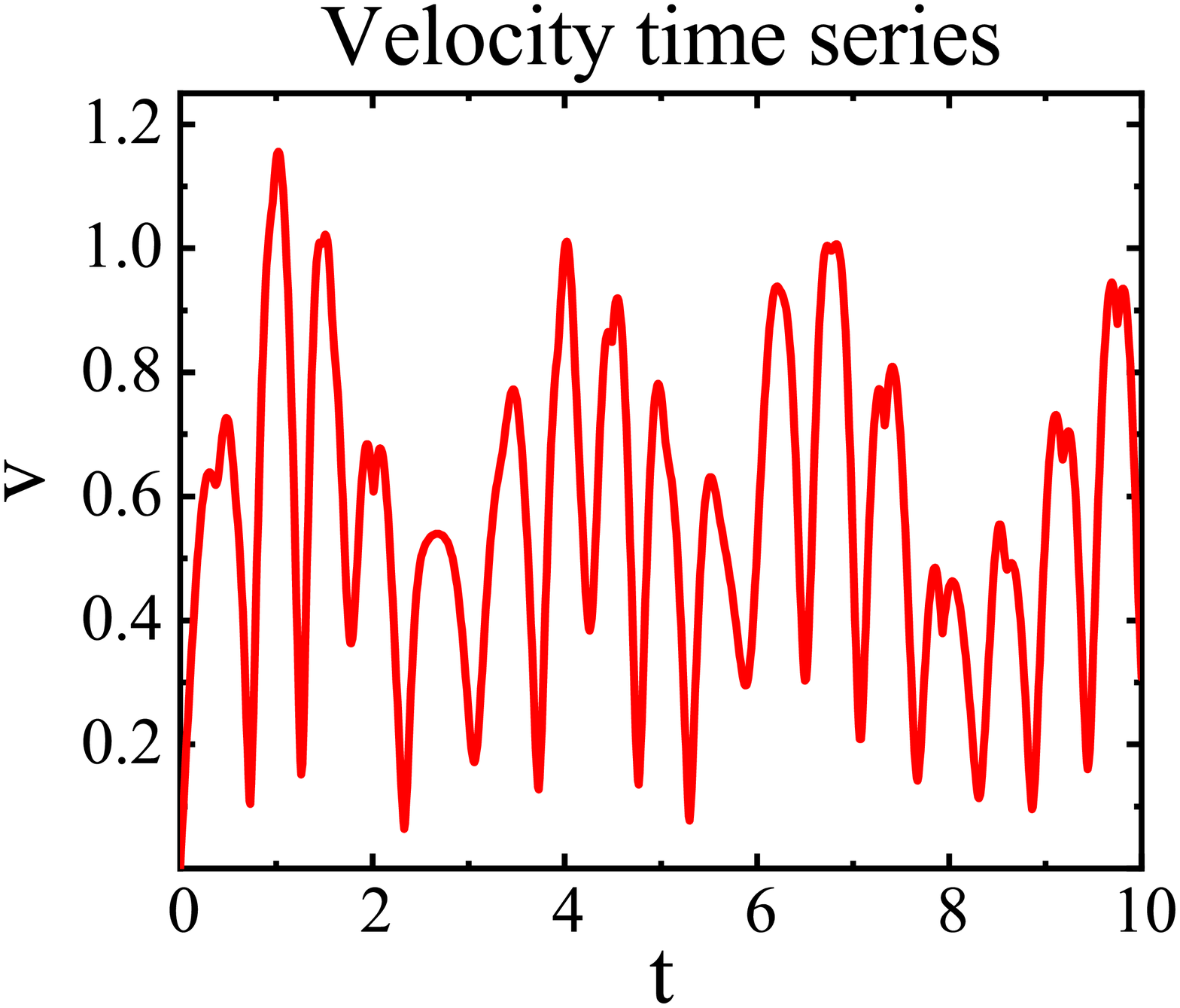}
		\end{minipage}
		\qquad
		\begin{minipage}{0.48\linewidth}
			\centering
			\includegraphics[width=1.0\linewidth]{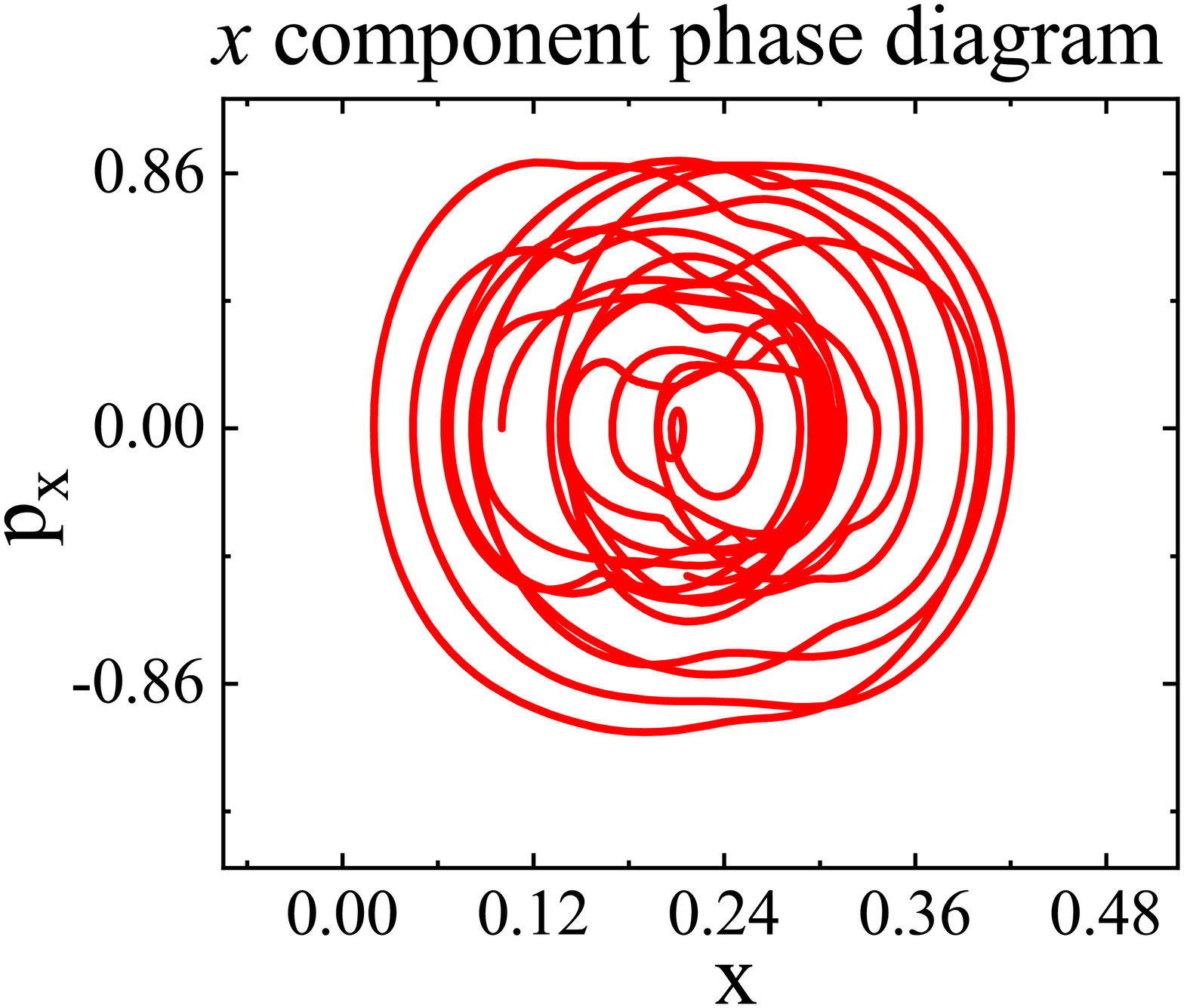}
		\end{minipage}
		\begin{minipage}{0.48\linewidth}
			\centering
			\includegraphics[width=1.0\linewidth]{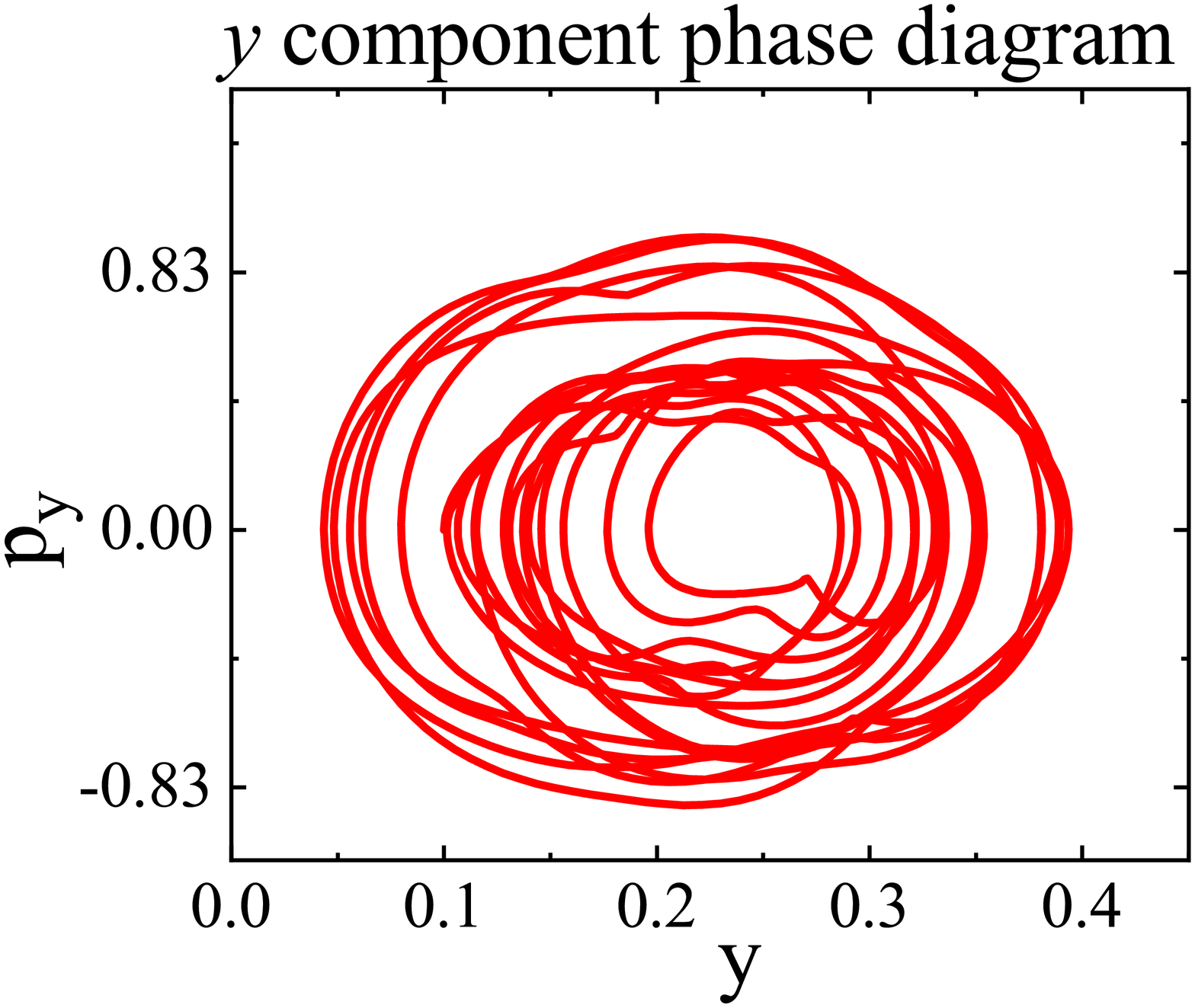}
		\end{minipage}
		\caption{{The state of motion of the first particle in a closed-loop system.}}
		\label{fig:j3}
	\end{figure}

	\emph{step 3:} the closed-loop system is restored according to the original initial position of the particle. Set the initial conditions and use the computer to solve the system of equations. The result is shown in Fig. \ref{fig:j3}.
	
\end{document}